\documentstyle[12pt,epsfig]{article}

\catcode`\@=11
\def\@citex[#1]#2{%
\if@filesw \immediate \write \@auxout {\string \citation {#2}}\fi
\@tempcntb\m@ne \let\@h@ld\relax \def\@citea{}%
\@cite{%
  \@for \@citeb:=#2\do {%
    \@ifundefined {b@\@citeb}%
      {\@h@ld\@citea\@tempcntb\m@ne{\bf ?}%
      \@warning {Citation `\@citeb ' on page \thepage \space undefined}}%
      {\@tempcnta\@tempcntb \advance\@tempcnta\@ne%
      \@tempcntb\number\csname b@\@citeb \endcsname \relax%
      \ifnum\@tempcnta=\@tempcntb 
        \ifx\@h@ld\relax%
          \edef \@h@ld{\@citea\csname b@\@citeb\endcsname}%
        \else%
          \edef\@h@ld{\ifmmode{-}\else--\fi\csname b@\@citeb\endcsname}%
        \fi%
      \else
        \@h@ld\@citea\csname b@\@citeb \endcsname%
        \let\@h@ld\relax%
      \fi}%
    \def\@citea{,\penalty\@highpenalty\,}%
  }\@h@ld
}{#1}}

\def\@citeb#1#2{{[#1]\if@tempswa , #2\fi}}
\def\@citeu#1#2{{$^{#1}$\if@tempswa , #2\fi }}
\def\@citep#1#2{{#1\if@tempswa , #2\fi}}

\def\bcites{         
        \catcode`\@=11
        \let\@cite=\@citeb
        \catcode`\@=12
}

\def\upcites{         
        \catcode`\@=11
        \let\@cite=\@citeu
        \catcode`\@=12
}

\def\plaincites{      
        \catcode`\@=11
        \let\@cite=\@citep
        \catcode`\@=12
}

\newcount\hour
\newcount\minute
\newtoks\amorpm
\hour=\time\divide\hour by 60
\minute=\time{\multiply\hour by 60 \global\advance\minute by-\hour}
\edef\standardtime{{\ifnum\hour<12 \global\amorpm={am}%
        \else\global\amorpm={pm}\advance\hour by-12 \fi
        \ifnum\hour=0 \hour=12 \fi
        \number\hour:\ifnum\minute<10 0\fi\number\minute\the\amorpm}}
\edef\militarytime{\number\hour:\ifnum\minute<10 0\fi\number\minute}

\def\draftlabel#1{{\@bsphack\if@filesw {\let\thepage\relax
   \xdef\@gtempa{\write\@auxout{\string
      \newlabel{#1}{{\@currentlabel}{\thepage}}}}}\@gtempa
   \if@nobreak \ifvmode\nobreak\fi\fi\fi\@esphack}
        \gdef\@eqnlabel{#1}}
\def\@eqnlabel{}
\def\@vacuum{}
\def\marginnote#1{}
\def\draftmarginnote#1{\marginpar{\raggedright\scriptsize\tt#1}}
\def\draft{
        \pagestyle{plain}
        \overfullrule=2pt
        \oddsidemargin -.5truein
        \def\@oddhead{\sl \phantom{\today\quad\militarytime} \hfil
        \smash{\Large\sl DRAFT} \hfil \today\quad\militarytime}
        \let\@evenhead\@oddhead
        \let\label=\draftlabel
        \let\marginnote=\draftmarginnote
        \def\ps@empty{\let\@mkboth\@gobbletwo
        \def\@oddfoot{\hfil \smash{\Large\sl DRAFT} \hfil}
        \let\@evenfoot\@oddhead}
        \def\@eqnnum{(\theequation)\rlap{\kern\marginparsep\tt\@eqnlabel}%
        \global\let\@eqnlabel\@vacuum}  }

\def\blackfonts{
        \font\blackboard=msbm10 scaled\magstep1
        \font\blackboards=msbm8
        \font\blackboardss=msbm6
}

\def\nblack{            
        \def\ZZ{{Z \n{10} Z}}
        \def\NN{{N \n{14} N}}
        \def\CC{{C \n{11} C}}
        \def\RR{{R \n{11} R}}
        \def\QQ{{Q \n{12} Q}}
        \def\PP{{P \n{11} P}}
}

\def\prep{         
        \catcode`\@=11
        \input art10.sty
        \catcode`\@=12
        
        \let\small\null
        \def\blackfonts{
                \font\blackboard=msbm10
                \font\blackboards=msbm7
                \font\blackboardss=msbm5
        }
        \let\sl\it
        \twocolumn
        \sloppy
        \voffset=-2.54truecm
        \hoffset=-2.54truecm
        \flushbottom
        \parindent 1em
        \leftmargini 2em
        \leftmarginv .5em
        \leftmarginvi .5em
        \marginparwidth 48pt
        \marginparsep 10pt
        \setlength{\columnsep}{2truecm}
        \setlength{\textwidth}{25.4truecm}
        \setlength{\textheight}{17truecm}
        \baselineskip=16pt
        \oddsidemargin .18truein
        \evensidemargin .17truein
}

\def\eqalign#1{\null\,\vcenter{\openup\jot\m@th
  \ialign{\strut\hfil$\displaystyle{##}$&$\displaystyle{{}##}$\hfil
      \crcr#1\crcr}}\,}
\def\eqalignno#1{\displ@y \tabskip\centering
  \halign to\displaywidth{\hfil$\@lign\displaystyle{##}$\tabskip\z@skip
    &$\@lign\displaystyle{{}##}$\hfil\tabskip\centering
    &\llap{$\@lign##$}\tabskip\z@skip\crcr
    #1\crcr}}

\def\section{\@startsection {section}{1}{\z@}{3.ex plus 1ex minus
 .2ex}{2.ex plus .2ex}{\large\bf}}
\def\subsection{\@startsection{subsection}{2}{\z@}{2.75ex plus 1ex minus
 .2ex}{1.5ex plus .2ex}{\bf}}        

\def\appendix{{\newpage\section*{Appendix}}\let\appendix\section%
        {\setcounter{section}{0}
        \gdef\thesection{\Alph{section}}}\section}

\def\abstract{\if@twocolumn
\section*{Abstract}
\else 
\begin{center}
{\bf Abstract\vspace{-.5em}\vspace{0pt}}
\end{center}
\quotation
\fi}

\catcode`\@=12

\newcommand{\beq}{\begin{equation}}
\newcommand{\eeq}{\end{equation}}
\newcommand{\beqa}{\begin{eqnarray}}
\newcommand{\eeqa}{\end{eqnarray}}

\def\noj#1,#2,{{\bf #1} (19#2)\ }
\def\jou#1,#2,#3,{{\sl #1\/ }{\bf #2} (19#3)\ }
\def\ann#1,#2,{{\sl Ann.\ Physics\/ }{\bf #1} (19#2)\ }
\def\cmp#1,#2,{{\sl Comm.\ Math.\ Phys.\/ }{\bf #1} (19#2)\ }
\def\ma#1,#2,{{\sl Math.\ Ann.\/ }{\bf #1} (19#2)\ }
\def\ng#1,#2,{{\sl Nagoya.\ Math.\ J.\/ }{\bf #1} (19#2)\ }
\def\jd#1,#2,{{\sl J.\ Diff.\ Geom.\/ }{\bf #1} (19#2)\ }
\def\invm#1,#2,{{\sl Invent.\ Math.\/ }{\bf #1} (19#2)\ }
\def\cq#1,#2,{{\sl Class.\ Quantum Grav.\/ }{\bf #1} (19#2)\ }
\def\cqg#1,#2,{{\sl Class.\ Quantum Grav.\/ }{\bf #1} (19#2)\ }
\def\ijmp#1,#2,{{\sl Int.\ J.\ Mod.\ Phys.\/ }{\bf A#1} (19#2)\ }
\def\jmphy#1,#2,{{\sl J.\ Geom.\ Phys.\/ }{\bf #1} (19#2)\ }
\def\jams#1,#2,{{\sl J.\ Amer.\ Math.\ Soc.\/ }{\bf #1} (19#2)\ }
\def\grg#1,#2,{{\sl Gen.\ Rel.\ Grav.\/ }{\bf #1} (19#2)\ }
\def\mpl#1,#2,{{\sl Mod.\ Phys.\ Lett.\/ }{\bf A#1} (19#2)\ }
\def\nc#1,#2,{{\sl Nuovo Cim.\/ }{\bf #1} (19#2)\ }
\def\np#1,#2,{{\sl Nucl.\ Phys.\/ }{\bf B#1} (19#2)\ }
\def\pl#1,#2,{{\sl Phys.\ Lett.\/ }{\bf #1B} (19#2)\ }
\def\pla#1,#2,{{\sl Phys.\ Lett.\/ }{\bf #1A} (19#2)\ }
\def\pr#1,#2,{{\sl Phys.\ Rev.\/ }{\bf #1} (19#2)\ }
\def\prd#1,#2,{{\sl Phys.\ Rev.\/ }{\bf D#1} (19#2)\ }
\def\prl#1,#2,{{\sl Phys.\ Rev.\ Lett.\/ }{\bf #1} (19#2)\ }
\def\prp#1,#2,{{\sl Phys.\ Rept.\/ }{\bf #1C} (19#2)\ }
\def\ptp#1,#2,{{\sl Prog.\ Theor.\ Phys.\/ }{\bf #1} (19#2)\ }
\def\ptpsup#1,#2,{{\sl Prog.\ Theor.\ Phys.\/ Suppl.\/ }{\bf #1} (19#2)\ }
\def\rmp#1,#2,{{\sl Rev.\ Mod.\ Phys.\/ }{\bf #1} (19#2)\ }
\def\yadfiz#1,#2,#3[#4,#5]{{\sl Yad.\ Fiz.\/ }{\bf #1} (19#2) #3%
\ [{\sl Sov.\ J.\ Nucl.\ Phys.\/ }{\bf #4} (19#2) #5]}
\def\zh#1,#2,#3[#4,#5]{{\sl Zh.\ Exp.\ Theor.\ Fiz.\/ }{\bf #1} (19#2) #3%
\ [{\sl Sov.\ Phys.\ JETP\/ }{\bf #4} (19#2) #5]}

\def\beq{\begin{equation}}
\def\eeq{\end{equation}}
\def\beqar{\begin{eqnarray}}
\def\eeqar{\end{eqnarray}}

\newcommand{\be}{\begin{equation}}
\newcommand{\ee}{\end{equation}}
\newcommand{\bea}{\begin{eqnarray}}
\newcommand{\eea}{\end{eqnarray}}

\def\nfrac#1#2{{\displaystyle{\vphantom1\smash{\lower.5ex\hbox{\small$#1$}}%
        \over\vphantom1\smash{\raise.25ex\hbox{\small$#2$}}}}}

\def\n#1{\mskip-#1mu}

\def\lae{\mathrel{\mathop{\smash{\lower .5 ex \hbox{$\stackrel<\sim$}}}}}
\def\lae{\mathrel{\mathop{\smash{\lower .5 ex \hbox{$\stackrel>\sim$}}}}}

\def\tr{{\rm Tr}}
\def\l:{\mathopen{:}\,}
\def\r:{\,\mathclose{:}}

\catcode`\@=11
\def\theequation{\arabic{equation}}
\catcode`\@=12

\nblack
\bcites

\nblack

\catcode`\@=11
\def\theequation{\thesection.\arabic{equation}}
\@addtoreset{equation}{section}
\catcode`\@=12

\newcommand{\beqn}{\begin{equation}}
\newcommand{\eeqn}{\end{equation}}
\newcommand{\beqnarray}{\begin{eqnarray}}
\newcommand{\eeqnarray}{\end{eqnarray}}
\newcommand {\bear} [1] {\begin {array} {#1}}
\newcommand {\ear} {\end {array}}

\newcommand {\beqarn} {\begin{eqnarray*}}
\newcommand {\eeqarn} {\end{eqnarray*}}

\begin{document}

\begin{titlepage}

\begin{center}
\today
\hfill LBNL-42655, UCB-PTH-98/??\\
\hfill                  hep-th/9812240

\vskip 1.5 cm
{\large \bf Large $N$ Elliptic Genus and AdS/CFT Correspondence}\\
\vskip 1 cm 
{Jan de Boer}\\
\vskip 0.5cm
{\sl Department of Physics,
University of California at Berkeley\\
366 Le\thinspace Conte Hall, Berkeley, CA 94720-7300, U.S.A.\\
and\\
Theoretical Physics Group, Mail Stop 50A--5101\\
Ernest Orlando Lawrence Berkeley National Laboratory\\
Berkeley, CA 94720, U.S.A.\\}

\end{center}

\vskip 0.5 cm

\begin{abstract}

According to one of Maldacena's dualities,
type IIB string theory on $AdS_3 \times S^3 \times K3$ is 
equivalent to a certain $N=(4,4)$ superconformal field theory.
In this note we compute the elliptic genus of the boundary
theory in the supergravity approximation. A finite quantity
is obtained once we introduce a particular exclusion
principle. In the regime where the supergravity approximation
is reliable, we find exact agreement with the elliptic genus
of a sigma model with target space $K3^N/S_N$.

\end{abstract}

\end{titlepage}

\section{Introduction}

One of the prime examples of the AdS$\leftrightarrow$CFT
duality of \cite{mal} is the duality between  
IIB string theory on $M^4\times S^3 \times AdS_3$
and certain two-dimensional conformal field theories.
This duality has its origin in a system of parallel
D1 and D5 branes, where the D5 branes are wrapped on some
four manifold $M^4$ which can be either $T^4$ or $K3$.   
The two-dimensional conformal field theory in question is the
infrared limit of the Higgs branch of the
gauge theory on the D1-branes. It has been argued
to be a deformation of the
$N=(4,4)$ sigma model with target 
space $(M^4)^N/S_N$ \cite{va1,vast}.   

There are several interesting issues regarding the precise
nature of the sigma model target space and the correspondence
between the sigma model moduli and the IIB string theory moduli.
These were recently discussed in \cite{dijkgraaf}, where (for $M=K3$)
it was shown that the sigma model has as target space 
the singular orbifold $K3^N/S_N$ with vanishing world-sheet
theta angle at the singularities. 
To resolve the metric singularities, one needs to turn
on a self-dual NS two-form on $K3$, to change the world-sheet
theta angle, one needs to turn on the RR zero and four form
field strengths on $K3$. A vanishing world-sheet theta angle
was also encountered (for $M={\bf R}^4$) in \cite{wit3}. 
Unfortunately, theories with theta angle equal to zero are
presumably ill-behaved, and differ considerably from conventional
conformal field theory orbifolds, which have theta angle
equal to $\pi$ \cite{aspinwall}. 
Several phenomena are probably closely related to the
vanishing of the world-sheet theta angle. One of these is
that although according to \cite{wit3} the Coulomb and Higgs
branches of the D1/D5 gauge theory decouple, the Higgs branch
remains connected to a separate branch coming from a 
neighborhood of the origin in moduli space \cite{throat}.
This region corresponds to the throat region
of the black hole description of the near extremal NS fivebrane.
Another phenomenon is the possibility for sufficiently excited
AdS$_3$ to decay by emitting branes \cite{frag}. Both phenomena
seem to disappear once we turn on RR zero- and four-form field
strengths.

As long as there is no phase transition associated with the
vanishing of the world-sheet theta angle, it should be possible
to compare BPS quantities obtained from supergravity
to those computed at the conformal field theory orbifold
point. Previously we showed that Kaluza-Klein spectrum of
supergravity reproduces the set of chiral primaries of the
$K3^N/S_N$ conformal field theory for large $N$ \cite{prev}.
In this note we extend this result and show that there is also
an exact agreement on the level of the elliptic genus. 
Some numerical evidence for this was given in \cite{prev}.
Since the
elliptic genus of $K3^N/S_N$ grows linearly with $N$ 
as $N \rightarrow \infty$, we cannot compare the supergravity
elliptic genus to that of the $N=\infty$ conformal field theory
(both are infinite). In order to make a meaningful comparison,
we have to introduce an ``exclusion principle'' 
\cite{mastr} on the
supergravity side, so that we can make a comparison for
finite and large $N$. We will determine such an
exclusion principle by requiring that the corresponding
truncated supergravity spectrum yields the precise set
of chiral primaries of $K3^N/S_N$ for finite $N$. 
This will be explained in section~2. In sections~3,4 and~5 we 
use the exclusion principle to compute the elliptic
genus for supergravity, and shows that it agrees with that
of the orbifold CFT for states with conformal weight
$h$ satisfying $h\leq (N+1)/4$. This includes the regime
where supergravity is reliable, but the agreement is valid
beyond the supergravity regime until black holes in $AdS_3$ 
start to form. Some final comments are given in section~6.

\section{An Exclusion Principle}

In \cite{prev} the Kaluza-Klein spectrum of six-dimensional
$N=(2,0)$ supergravity compactified on $AdS_3 \times S^3$
was computed (see also \cite{sez1,larsen}). This six-dimensional
supergravity is obtained by compactifying type IIB string theory
on $AdS_3 \times S^3$, and because the size of the $K3$ is
much smaller than that of $ AdS_3$ and $S^3$ the Kaluza-Klein
spectrum of six-dimensional supergravity contains the lightest
states of type IIB string theory on $ AdS_3 \times S^3 \times K3$.
The resulting KK spectrum can be organized in representations
of the relevant $AdS$ supergroup, which is $SU(1,1|2)_L
\times SU(1,1|2)_R$. The group $SU(1,1|2)$ is generated
by the global modes $\{L_{\pm 1},L_0,G^i_{\pm 1/2},J^i_0\}$
of the $N=4$ superconformal algebra, and the bosonic part
of $SU(1,1|2)_L \times SU(1,1|2)_R$ is just the isometry
group $SO(2,2) \times SO(4)$ of $AdS_3 \times S^3$.
The supergroup $SU(1,1|2)$ has long and short representations,
but in the KK spectrum of supergravity only short representations
appear. Short representations of $SU(1,1|2)$ will be denoted by
$(j)_S$, and short representations of $SU(1,1|2)_L \times SU(1,1|2)_R$
by $(j,j')_S$. A short representation $(j)_S$ is obtained by taking
as highest weight state a chiral primary of the $N=4$ superconformal algebra $|h,j\rangle$ that satisfies $L_0|h,j\rangle = h|h,j\rangle$, $J_0^3|h,j\rangle=
j|h,j\rangle$ with $h=j/2$, and by acting on it with 
$\{L_{\pm 1},L_0,G^i_{\pm 1/2},J^i_0\}$. The quantum numbers
of the states in such a short representation are 
\be \label{quantumno}
\begin{array}{c|l}
{\rm multiplicity} & {\rm states}  \\ \hline
1 & |\frac{j}{2} + k,j\rangle, 
|\frac{j}{2} + k,j-2\rangle, \ldots, |\frac{j}{2} + k,-j\rangle,\,\, k\geq 0 
\\ 
2 & |\frac{j+1}{2} + k,j-1\rangle, 
|\frac{j+1}{2} + k,j-3\rangle, 
\ldots, |\frac{j+1}{2} + k,1-j\rangle,\,\, k\geq 0 \\
1 & |\frac{j+2}{2} + k,j-2\rangle, 
|\frac{j+2}{2} + k,j-4\rangle, \ldots,
 |\frac{j+2}{2} + k,2-j\rangle,\,\, k\geq 0 
\end{array}
\ee
where $k$ is an arbitrary nonnegative integer that appears because of the freedom
to act $k$ times with $L_{-1}$ on each state. Notice that each short
representation contains precisely one chiral primary.
The KK spectrum of
six-dimensional supergravity can be organized in the following short
representations \cite{prev}
\bea
&& (0,2)_S, (1,3)_S,(2,4)_S, \ldots, \nonumber \\
&& 21(1,1)_S, 22(2,2)_S, 22(3,3)_S,\ldots, \nonumber \\
&& (2,0)_S, (3,1)_S, (4,2)_S, \ldots
\label{kktable}
\eea
The representations $(0,2)_S$ and $(2,0)_S$ contain the higher
modes $L_{-(2+k)},G^i_{-(3/2+k)},J^i_{-(1+k)}$, $k\geq 0$,
of the $N=4$ superconformal algebra,
as one easily sees from (\ref{quantumno}). These representations
do not correspond to propagating degrees of freedom in the bulk,
but to degrees of freedom living only on the boundary (sometimes
called ``singletons''). For the Virasoro generators this was
shown in \cite{hencs} by relating them to a particular class of
diffeomorphisms, and one can easily generalize this to the case
with supersymmetry. 

The set of (left $\times$ right) chiral primaries of
an $N=4$ superconformal field theory 
can be conveniently encoded in terms of the generalized
Poincar\'e polynomial
\be
P_{t,\bar{t}}  = \tr( t^{J_0} \bar{t}^{\bar{J}_0} )
\ee
where the trace is taken over the space of chiral primaries only.
In case the superconformal field theory is a supersymmetric sigma
model with target space $M$, the Poincar\'e polynomial equals
\be
P_{t,\bar{t}} = \sum_{p,q} h_{p,q} t^p \bar{t}^q
\ee
where $h^{p,q}$ are the Betti numbers of $M$ \cite{witten1}.
The Poincar\'e polynomial of a resolution of $K3^N/S_N$ called the
Hilbert scheme of $N$ points on $K3$
was computed in \cite{goso} and has generating function
\bea \label{Poincare}
\sum_{N\geq0} p^N
P_{t,\bar{t}}(K3^N/S_N) & = & 
\prod_{m=1}^{\infty} (1-p^m t^{m-1} \bar{t}^{m-1})^{-1}
 (1-p^m t^{m-1} \bar{t}^{m+1})^{-1}
 (1-p^m t^{m+1} \bar{t}^{m-1})^{-1} \nonumber \\
& & \qquad \times 
 (1-p^m t^{m+1} \bar{t}^{m+1})^{-1}
 (1-p^m t^{m} \bar{t}^{m})^{-20} .
\eea

We have to compare (\ref{kktable}) to the set of chiral primaries
of the $K3^N/S_N$ SCFT for large $N$.
It is easy to see that $P_{t,\bar{t}}(K3^N/S_N)$
does have a well-defined $N\rightarrow \infty$
limit. Because (\ref{Poincare})
has a single factor of $(1-p)^{-1}$, it
is of the form
\be
a_0  + (a_0 + a_1)p + (a_0+a_1+a_2) p^2 + \ldots =
(1-p)^{-1} (a_0 + a_1 p + a_2 p^2 + \ldots ) .
\ee
Thus the $N\rightarrow \infty$ limit is obtained by extracting the factor
of $(1-p)^{-1}$ and taking  $p\rightarrow 1$, which yields
\be \label{e56}
P_{t,\bar{t}}(K3^{\infty}/S_{\infty}) =
(1-t\bar{t})^{-21} \prod_{m=1}^{\infty}
(1- t^{m-1} \bar{t}^{m+1})^{-1}
(1- t^{m+1} \bar{t}^{m-1})^{-1}
(1- t^{m+1} \bar{t}^{m+1})^{-22} .
\ee
Since there is a factor $(1-t^j \bar{t}^{j'})^{-1}$ 
in (\ref{e56}) for each
short multiplet $(j,j')_S$ in (\ref{kktable}), the multiparticle
chiral primaries of supergravity are in one to one correspondence
with those of a sigma model on $K3^{\infty}/S_{\infty}$ \cite{prev}.

It turns out that we can also get an agreement for finite $N$, once
we introduce a suitable exclusion principle. Each of the factors
$(1-t^j \bar{t}^{j'})^{-1}$ in (\ref{e56}) came from a factor
$(1- p^{d(j,j')} t^j \bar{t}^{j'})^{-1}$ in (\ref{Poincare}). If follows
then from (\ref{Poincare}) that if we associate degree 
$d(j,j')$ to the chiral 
primary in $(j,j')_S$ and keep only those products of chiral primaries
which have total degree $\leq N$, we recover precisely the set of chiral
primaries of $K3^N/S_N$ for finite $N$. We will write
$(j,j',d)_S$ for a short multiplet to indicate the degree $d$ of the
corresponding chiral primary. By comparing (\ref{e56}) and
(\ref{Poincare}), we find that the table of short multiplets including
their degrees reads 
\bea
(m,m+2;m+1)_S && m=0,1,2,\ldots \nonumber \\
(m,m;m-1)_S && m=2,3,4,\ldots   \nonumber \\
20 (m,m;m)_S && m=1,2,3,\ldots   \nonumber \\
(m,m;m+1)_S && m=1,2,3,\ldots   \nonumber \\
(m+2,m;m+1)_S && m=0,1,2,\ldots   . 
\label{kktable2}
\eea

Although so far we only associated a degree to the chiral primaries,
we propose to associate the same degree to all descendants of the
chiral primary. At this moment, we don't have a particularly good
reason to do so. However, it will turn out that if we extend
the exclusion principle to include all descendants of 
chiral primaries, and keep only products of states
whose total degree is $\leq N$,
we also find precise agreement for the
finite $N$ elliptic genus. Thus, to summarize, for the
finite $N$ supergravity Hilbert space we propose
the following direct sum of tensor 
products\footnote{The tensor product of two short representations
of $SU(1,1|2)$ contains one short and many long representations.
Therefore, the Hilbert space (\ref{Hilbert}) contains many
states whose scaling dimensions are not protected by supersymmetry.}
of
$SU(1,1|2)_L \times SU(1,1|2)_R$ representations,
\be \label{Hilbert}
{\cal H}^{(N)}_{\rm sugra} = \bigoplus_{\stackrel{\{j_i,j'_i;d_i\}}{
\sum d_i \leq N}} \bigotimes_i (j_i,j'_i;d_i)_S .
\ee
It would be interesting to have some independent evidence for
the exclusion principle  $\sum d_i \leq N$, for instance
from conformal field theory. 

In the multiparticle Hilbert space (\ref{Hilbert}), the generators
of the Virasoro algebra appear in a quite asymmetric way. The
generators $L_{\pm 1},L_0$ are part of $SU(1,1|2)$ and 
their action does not change the particle number, whereas 
$L_{-2},L_{-3},\ldots$ appear in a separate $SU(1,1|2)$
representation and their action does change the particle number.
In particular, for two chiral primaries $A$ and $B$, the two
states $(L_{-1} A)B$ and $A(L_{-1} B)$ appear as two inequivalent
two-particle states in (\ref{Hilbert}), whereas $(L_{-2} A)B$ 
and $A(L_{-2} B)$ appear as two equivalent three-particle states. 
This is in agreement with conformal field theory, since in conformal
field theory $(L_{-1} A)B$ and $A(L_{-1} B)$ are independent, whereas
the difference of $(L_{-2} A)B$ and $A(L_{-2} B)$ is proportional 
to the two-particle state $L_{-1}^2(AB)$. Therefore,
$(L_{-2} A)B$ and $A(L_{-2} B)$ should not be counted independently. 
Another feature of (\ref{Hilbert}) is that arbitrary high
powers of $L_{-1}$ appear in it, whereas the power of $L_{-2}$ can
never be larger than $N$.

\section{Large $N$ Elliptic Genus of $K3^N/S_N$}
 
The spectrum of left and right-moving chiral primaries is not
the only part of the spectrum which is independent of marginal
deformations of the theory. A more general object with this
property is the elliptic genus, which can only change if a phase
transition occurs. The elliptic genus is defined by
\be \label{defellgen}
Z(\tau,z)=\tr_{RR} (-1)^F q^{L_0-c/24} \bar{q}^{\bar{L}_0-c/24} y^{J^3_0}
\ee
with $q=e^{2\pi i\tau}$ and $y=e^{2\pi i z}$, and the trace is over
the Ramond sector of the Hilbert space \cite{ell1,ell2,ell3}.
The elliptic genus receives only contributions from states of
the form $|{\rm anything}\rangle_L \otimes |{\rm groundstate}\rangle_R$, 
which are $1/4$ BPS states of the conformal field theory. 

The elliptic genus for $K3$ equals \cite{ell4,ell5}
\be
\label{ellgenus}
Z(\tau,z) \equiv\sum_{m,l} c(m,l) q^m y^l= 24 \left(
\frac{\theta_3(\tau,z)}{\theta_3(\tau,0)}\right)^2 - 
2 \frac{\theta_4(\tau,0)^4 - \theta_2(\tau,0)^4}{\eta(\tau)^4}
\left(\frac{\theta_1(\tau,z)}{\eta(\tau)} \right)^2.
\ee
With this definition of $c(m,l)$, the elliptic genus of $K3^N/S_N$ 
has generating function \cite{dmvv}
\be \label{ellgen2}
\sum_{N\geq 0}  p^N Z(K3^N/S_N;\tau,z) = 
\prod_{n>0,m\geq0,l} \frac{1}{(1-p^n q^m y^l)^{c(nm,l)}}.
\ee
The coefficients $c(m,l)$ are a function of $4m-l^2$ only,
\be \label{aa1}
c(m,l) \equiv c(4m-l^2),
\ee
and the first few values are
$c(r)=0, r<-1$, $c(-1)=2$, $c(0)=20$. In contrast to the Poincar\'e
polynomial considered in the previous section, the elliptic genus
of $K3^N/S_N$ is not finite in the $N\rightarrow \infty$ limit,
but diverges linearly as $N\rightarrow \infty$. The origin of
this linear divergence is the presence of a factor $1/(1-p)^2$ in
(\ref{ellgen2}). Multiplying $1/(1-p)^2$ with a finite power
series in $p$ gives a series whose coefficients diverge linearly
for large $N$,
\be \label{auxid}
(1-p)^{-2} (a_0  + a_1 p + a_2 p^2 + \ldots) = \ldots + 
((N+1)a_0 + Na_1 + (N-1)a_2 + \ldots)p^N + \ldots  .
\ee
 Because of this divergence, we cannot simply compare the
supergravity result to the conformal field theory result for $N=\infty$. 
However, thanks to the exclusion principle described above, we
can compare the elliptic genus for finite $N$. When doing so,
we should keep in mind that the supergravity
states live in the NS sector of the theory, whereas the elliptic genus
(\ref{defellgen}) is defined as a trace over the RR sector. 
It will therefore be convenient to work with an analogue of 
the elliptic genus defined directly in the NS sector. 
Since spectral flow establishes an isomorphism between RR states of
the form $|{\rm anything}\rangle_L \otimes |{\rm groundstate}\rangle_R$
and NS states of the form 
$|{\rm anything}\rangle_L \otimes |{\rm chiral\,\,primary}\rangle_R$,
we can define an NS elliptic genus via
\be \label{defnsellgen}
Z^{\rm cft}_{NS}(q,y)=\tr_{
|{\rm anything}\rangle_L \otimes |{\rm chiral\,\,primary}\rangle_R} (-1)^F q^{L_0}  y^{J^3_0} .
\ee
Spectral flow maps a Ramond
 state with conformal weight $h_R$ and $J^3_0$ eigenvalue $q_R$
to an NS state with conformal weight $h_{NS}=
c/24 + h_R+q_R/2$ and $J^3_0$ eigenvalue
$q_{NS}=q_R+c/6$. It is then straightforward to show that 
\be \label{ellgen3}
\sum_{N\geq 0}  p^N Z^{\rm cft}_{NS}(K3^N/S_N;q,y) = 
\prod_{n>0,m,l} \frac{1}{(1-p^n q^m y^l)^{c_{\rm cft}(n,m,l)}}.
\ee
where $c_{\rm cft}(n,m,l)$ is related to $c(n)$ in (\ref{aa1}) via 
\be \label{defccft}
c_{\rm cft}(n,m,l)=c(4mn-n^2-l^2) 
\ee
and the product in (\ref{ellgen3}) 
is over $m,l$ that satisfy $2m\in{\bf Z}_{\geq 0}$,
$m-l/2 \in {\bf Z}_{\geq 0}$ and $2m \geq |l|$.

Our goal is to compare (\ref{ellgen3}) to the NS elliptic genus computed from
supergravity. However, we should not expect to find a complete
agreement, since supergravity only gives an appropriate description of
the spectrum for sufficiently low conformal weights, before string states
start to contribute\footnote{The first string states that
appear are Kaluza-Klein and winding string states
coming from the $K3$. Their left and right-moving conformal
weights $h$ and $\bar{h}$ satisfy $h+\bar{h}\sim (g_6 Q_5)^{1/2}$
and $h+\bar{h}\sim (g_6 Q_1)^{1/2}$ respectively.}.
We will see that there is only agreement if the conformal
weight of the states satisfies 
\be
\label{bound}
h \leq \frac{N+1}{4}.
\ee
This bound corresponds precisely to the point where the Ramond ground
state (which has $h=c/24=N/4$) appears. A particle with mass
$m$ that corresponds to a state with conformal weight $h=\bar{h}\sim
N/4$ introduces a deficit angle of $2\pi$ in the geometry, and
a black hole starts to form (see e.g. \cite{mart2}). Thus, the supergravity
approximation certainly breaks down for $h>N/4$. Interestingly,
the exclusion principle yields a proper description of all
$1/2$ BPS states with conformal weights up to $h\sim N/2$, which
overlaps with the black hole regime. On the other hand, it only
properly describes the $1/4$ BPS states for conformal weights up
to $h\sim N/4$. Apparently new $1/4$ BPS states but no new $1/2$
BPS states appear in the black hole phase.

The part of the elliptic genus that receives only contributions from states
with conformal weight satisfying (\ref{bound}) has a particularly
simple form.
Namely, we claim that
\bea \label{claim}
\sum_{N\geq 0}  p^N Z^{\rm cft}_{NS}(K3^N/S_N;q,y) & = &
\sum_{\stackrel{n,m,l}{m\leq (n+1)/4}}
 p^n q^m y^l (a_{\rm cft}(m,l) n + b_{\rm cft}(m,l)) \nonumber \\ & & +
\sum_{\stackrel{n,m,l}{m> (n+1)/4}} (\ldots) .
\eea
The claim follows from (\ref{auxid}), once we show that for given $m,l$
there is at most one $n$ that satisfies
$n/4 \geq m$ \footnote{That we have $n$ instead of $n+1$ in
this inequality is due to the following fact: if the left hand
side of (\ref{auxid}) truncates at $a_r$, then the coefficients
on the right hand side are linear functions of $n$ for $n\geq r-1$.}
and for which
$c_{\rm cft}(n,m,l)=c(4mn-n^2-l^2)\neq 0$. Since $c(r)=0$ for $r<-1$, 
the largest value of $n$ for which $c(4mn-n^2-l^2)\neq 0$ is
always $4m$, thereby proving the claim.

Altogether we have shown that the part of the elliptic genus that
can be meaningfully compared to supergravity is the part that behaves
as a linear function of $N$. What we show next is that the supergravity
elliptic genus has a decomposition exactly analogous to that in 
(\ref{claim}), with coefficients $a(m,l)$ and $b(m,l)$
identical to those
appearing in the conformal field theory elliptic genus.

\section{Large $N$ Elliptic Genus of Supergravity}

To define the supergravity elliptic genus we use exactly the
same definition as in conformal field theory, taking a trace
over the supergravity Hilbert space (\ref{Hilbert}) rather than
the conformal field theory Hilbert space. Since supergravity describes
the NS sector of the theory, we should use the definition of
the NS elliptic genus in (\ref{defnsellgen}). Thus, we define
\be \label{defnsellgen2}
Z^{{\rm sugra},{(N)}}_{NS}(q,y)=\tr_{
|{\rm anything}\rangle_L 
\otimes |{\rm chiral\,\,primary}\rangle_R 
\in {\cal H}^{(N)}_{\rm sugra}} (-1)^F q^{L_0}  y^{J^3_0} .
\ee
Due to the multiparticle form of the supergravity Hilbert space,
we can write the generating function for the elliptic genera
(\ref{defnsellgen2}) in a form similar to (\ref{ellgen3}),
\be \label{ellgen4}
\sum_{N\geq 0}  p^N Z^{{\rm sugra},{(N)}}_{NS}(q,y) = 
\prod_{n>0,m,l} \frac{1}{(1-p^n q^m y^l)^{c_{\rm sugra}(n,m,l)}}.
\ee
The powers $c_{\rm sugra}(n,m,l)$ appearing in (\ref{ellgen4})
are completely different from the powers $c_{\rm cft}(n,m,l)$
appearing in (\ref{ellgen3}). The number
$c_{\rm sugra}(n,m,l)$ is the number of single particle states of
degree $n$ in supergravity, counted with a sign $(-1)^F$,
of the form $|{\rm anything}\rangle_L
\otimes |{\rm chiral\,\,primary}\rangle_R$, 
where the conformal weight and $J_0^3$ eigenvalue of 
$|{\rm anything}\rangle_L$ are $m$ and $l$ 
respectively\footnote{Except for $c_{\rm sugra}(1,0,0)$, which is
equal to $2$.}.
It follows from (\ref{kktable2}) that $c_{\rm sugra}(n,m,l) \leq 48$.
On the other hand, the numbers
$c_{\rm cft}(n,m,l)$ grow exponentially. It is therefore
quite remarkable that there is any relation at all between
(\ref{ellgen3}) and (\ref{ellgen4}).

To demonstrate this relation, we first show that (\ref{ellgen4})
can be decomposed in the same way as in (\ref{claim}), namely
\bea \label{claim2}
\sum_{N\geq 0}  p^N Z^{{\rm sugra},{(N)}}_{NS}(q,y) & = & 
\sum_{\stackrel{n,m,l}{m\leq (n+1)/4}}
 p^n q^m y^l (a_{\rm sugra}(m,l) n + b_{\rm sugra}(m,l)) \nonumber \\ & & +
\sum_{\stackrel{n,m,l}{m> (n+1)/4}} (\ldots) .
\eea
The reasoning is similar to that below (\ref{ellgen3}). First,
we notice that (\ref{defnsellgen2}) has a factor of $1/(1-p)^2$,
just like (\ref{defnsellgen}), which is responsible for the linear
growth of the elliptic genus as a function of $N$. Next, by inspecting the
table of KK states (\ref{kktable2}) we see that for any given
$L_0$ eigenvalue $m$ and $J_0^{3}$ eigenvalue $l$, there is
at most one KK state with degree $d\geq 4m$. We can now repeat
the argument given below (\ref{claim}) to establish (\ref{claim2}).

\section{Equivalence of Elliptic Genera}

As we said before, we certainly cannot trust supergravity to give
a reliable description of states with conformal weight $h\geq (N+1)/4$.
Therefore, if the AdS$\leftrightarrow$CFT duality is correct,
we should only expect 
\be a_{\rm cft}(m,l)=a_{\rm sugra}(m,l), \qquad
    b_{\rm cft}(m,l)=b_{\rm sugra}(m,l) .
\label{req1}
\ee
These conditions are much weaker then the equivalence of the full
elliptic genera, which would imply $c_{\rm sugra}(n,m,l)=
c_{\rm cft}(n,m,l)$ and which is certainly not true.
The coefficients $a(m,l)$ and $b(m,l)$ are fixed by the residues
of (\ref{defnsellgen}) and (\ref{defnsellgen2}) at the
double pole $p=1$.
This can easily be seen from
(\ref{auxid}): If we write the left hand side of (\ref{auxid})
as $c_{-2}(1-p)^{-2} + c_{-1}(1-p)^{-1} + {\rm regular}$,
then $a=c_{-2}$ and $b=c_{-2}+c_{-1}$. These residues can
be extracted directly from (\ref{defnsellgen}) and (\ref{defnsellgen2}),
by removing the factor of $1/(1-p)^2$ and by evaluating the
remainder and its $p$-derivative at $p=1$. This leads to a useful
equivalent description of (\ref{req1}), namely we find that
(\ref{req1}) is valid if and only if the "first two moments"
of the $c_{\rm cft}$ and $c_{\rm sugra}$ agree, i.e.
\be \label{req2}
 \sum_n c_{\rm cft}(n,m,l) = \sum_n c_{\rm sugra}(n,m,l),
\qquad \sum_n n c_{\rm cft}(n,m,l) = \sum_n n c_{\rm sugra}(n,m,l) .
\ee
This clearly shows that the equivalence of the supergravity part
of the elliptic genera is much weaker than the equivalence of
the full elliptic genera, which would require the $c$'s themselves
to be identical.

In the remainder of this section, we compute the left and right
hand sides of (\ref{req2}) and show that they are the same.

We first consider $\sum_n c_{\rm cft}(n,m,l)$ and $\sum_n n 
c_{\rm cft}(n,m,l)$. The two crucial identities that allow
us to evaluate these sums are 
\be
\label{us1}
Z_{K3}(q,y)|_{y=1} = 24, \qquad \frac{\partial}{\partial y}
Z_{K3}(q,y)|_{y=1} = 0
\ee
which follow from (\ref{ellgenus}) and $\theta_1(\tau,0)=0$,
$\frac{\partial}{\partial z} \theta_3(\tau,z)|_{z=0} =0$. 
The identities (\ref{us1}) imply that the coefficients
$c(m,l)=c(4m-l^2)$ that appear in the expansion of (\ref{ellgenus})
obey
\bea
\label{us3}
\sum_l c(4m-l^2) & = & 24 \delta_{m,0} \\
\label{us2}
\sum_l l c(4m-l^2) & = & 0 .
\eea
Consider now for example $\sum_n c_{\rm cft}(n,m,l)$.
According to (\ref{defccft}) we get
\bea
\sum_{n>0} c_{\rm cft}(n,m,l) & = & \sum_{n>0} c(4nm-n^2-l^2) \nonumber \\
                          & = & \sum_{n>0} c(4u(t-u)-(n-t)^2)
\label{us4}
\eea
where in the second line we substituted $m=t/2$, $l=t-2u$, with $t,u$
nonnegative integers. The second line can be evaluated using (\ref{us3}),
where we have to be careful that the sum over $n$ in (\ref{us4}) may
not reproduce all terms in the sum over $l$ in (\ref{us3}) 
(this happens only for $|l|\leq 1$, in which case we also need to use
$c(-1)=2$ and $c(0)=20$).
In the same way, we can evaluate $\sum_{n>0} n c_{\rm cft}(n,m,l)$
using both (\ref{us3}) and (\ref{us2}). The results are summarized in
the following table
\be \label{resu1}
\begin{array}{c|c|c}
\mbox{{\rm values for $l$}}
& \sum_{n>0} c_{\rm cft}(n,m,l) & \sum_{n>0} n c_{\rm cft}(n,m,l) 
\\ \hline
|l|>1 & 24 \delta_{m,|l/2|} & 48 m \delta_{m,|l/2|} \\
|l|=1 & 24 \delta_{m,1/2}-2 & 48 m \delta_{m,1/2} \\
l=0 & 22 \delta_{m,0}-20 & 2 \delta_{m,0}  .
\end{array}
\ee

Next we turn to the sums $\sum_{n>0}c_{\rm sugra}(n,m,l)$ and
$\sum_{n>0} n c_{\rm sugra}(n,m,l)$. The easiest way to determine
these is by first computing the generating function
\be
s(p,q,y)= \sum_{n,m,l} c_{\rm sugra}(n,m,l) p^n q^m y^l
\ee
and by evaluating $s(p,q,y)$ and its $p$-derivative at $p=1$.
Since $c_{\rm sugra}(n,m,l)$ counts the number of single particle
KK states with particular quantum numbers and degrees, we can
construct $s(p,q,y)$ directly from table (\ref{kktable2}), and get
\bea \label{genfie}
(1-q)(y-y^{-1}) s(p,q,y) & = & 
2 p (1-q)(y-y^{-1}) \nonumber \\ & & 
 + (20p + p^2)(q^{1/2} (y^2-y^{-2}) - 2 q (y-y^{-1})) 
\nonumber \\ & & 
 + \frac{2p + 20 p^2 + 2 p^3}{1-pq^{1/2} y} (qy^3 - 2 q^{3/2} y^2 + q^2 y)  
\nonumber \\ & & 
 + \frac{2p + 20 p^2 + 2 p^3}{1-pq^{1/2} y^{-1}} (-qy^{-3} 
+ 2 q^{3/2} y^{-2} - q^2 y^{-1}) .
\eea
The factor of $(1-q)$ in the left hand side
has its origin in the fact that single
particle states can carry arbitrary powers of $L_{-1}$, 
the factor of $y-y^{-1}$ arises because $SU(2)$ representations of
spin $j/2$ give a contribution $(y^{j+1}-y^{-j-1})/(y-y^{-1})$.

From (\ref{genfie}) we obtain
\bea \label{resu2}
s(1,q,y) & = & 
\frac{-46+26 q - 2 q^{\frac{1}{2}} (y+y^{-1}) }{1-q} \nonumber \\
& & +
\frac{24}{1-q^{\frac{1}{2}} y} +
\frac{24}{1-q^{\frac{1}{2}} y^{-1}} \nonumber \\
\frac{\partial}{\partial p} s(p,q,y) |_{p=1} & = & 2+
\frac{24 q^{\frac{1}{2}} y}{(1-q^{\frac{1}{2}} y)^2} +
\frac{24 q^{\frac{1}{2}} y^{-1}}{(1-q^{\frac{1}{2}} y^{-1})^2} 
\eea

If we expand the results in (\ref{resu2}) we recover precisely
the results of (\ref{resu1}), which proves that the conformal
field theory and supergravity elliptic genus are equivalent in
the supergravity regime.

\section{Comments}

In this paper we have shown the equivalence  of the conformal
field theory and supergravity elliptic genus in
the supergravity regime, extending the 
numerical evidence in \cite{prev}. Although the definition of the 
supergravity elliptic genus involves one piece of information
which cannot be obtained purely within supergravity, namely
an exclusion principle, we consider the fact that the elliptic
genera agree as very strong evidence in favor of both the 
proposed exclusion principle as well as the 
AdS$\leftrightarrow$CFT correspondence. In addition, this
completely resolves the puzzle in \cite{vafap}.

Originally, one of the motivations for this work was
to examine to what extend supergravity can account for the
entropy of black holes. In \cite{vast} this entropy was
determined by looking at the degeneracy of $1/4$ BPS states,
which are counted by the elliptic genus. However, the region
of interest is the one where the conformal weight is much
greater than the central charge, $h \gg 6N$. This is not the
region $h\leq (N+1)/4$ for which we proved the equivalence.
The supergravity elliptic genus grows much slower for $h \gg 6N$
than the conformal field theory elliptic genus, because 
$c_{\rm sugra}(n,m,l)$ is bounded,
whereas $c_{\rm cft}(n,m,l)$ grows exponentially as a function
of $m$. 
Thus supergravity, subject to the exclusion principle, cannot
account for the entropy of the black hole. On the other
hand, if one drops the exclusion principle, the number of supergravity
states grows more rapidly than those of the conformal field theory.
As a function of energy, the six-dimensional 
supergravity entropy scales as
$E^{5/6}$ but the conformal field theory entropy only as $E^{1/2}$
\cite{hooo}. Therefore, a more generous exclusion principle
might allow one to encode the entropy of the black hole in
terms of supergravity states. However, since stringy states are
clearly important for the description of the black hole,
such a supergravity description would be a convenient 
parametrization rather than an adequate microscopic description
of the states that make up the black hole entropy.

Along these lines, it is intriguing to notice that equations
(\ref{req2}) have a unique solution for $c_{\rm cft}(n,m,l)$
in terms of $c_{\rm sugra}(n,m,l)$, once we know that
$c_{\rm cft}(n,m,l)$ is of the form $c(4mn-n^2-l^2)$. Thus,
it is possible to reconstruct the $K3^N/S_N$ elliptic
genus from the supergravity data, although it remains to 
be seen whether this has any physical meaning.

\noindent
{\bf Acknowledgement}

The author would like to thank the participants and organizers
of the Amsterdam
Summer Workshop ``String Theory and Black Holes'' and the
Aspen Summer Workshop ``$M$-Theory and Black Holes'' for
discussions and hospitality, and K.~Hori, S.~Kachru and 
J.~Maldacena for useful discussions.
This work was supported in part by the NSF
grant PHY-95-14797 and the DOE grant DE-AC03-76SF00098.

\end{document}